\documentclass[aps,prl,twocolumn,superscriptaddress,showpacs]{revtex4}
\begin{document}
\title{Trapping Quantum Coherence in Local Energy Minima}
\author{Haiqing Wei}
\email[]{davidhwei@yahoo.com}
\affiliation{Department of Physics, McGill University, Montreal, Quebec, Canada H3A 2T8}
\author{Xin Xue}
\affiliation{Department of Natural Resource Sciences, McGill University, Ste-Anne-de-Bellevue, Quebec, Canada H9X 3V9}
\author{S.D. Morgera}
\affiliation{Department of Electrical Engineering, McGill University, Montreal, Quebec, Canada H3A 2A7}
\begin{abstract}
Clusters of solid-state quantum devices have long-living metastable states of local energy minima which may be used to store quantum information. The low to vanishing rate of dissipation fulfils the prerequisite to maintain quantum coherence. Then physical symmetrization of the devices could minimize the couplings of the clusters to environmental degrees of freedom so to reduce the rate of decoherence. Combined with various other error correction mechanisms and methods, such designs and optimizations could render solid-state devices useful for quantum information processing, which have the advantages of flexibility in state manipulation and system scaling.
\end{abstract}
\maketitle

Quantum computers are entrancing machines being able to carry out exceedingly fast algorithms by virtue of the so-called quantum parallelism \cite{FeynmanEtAl}. Recent research has already showed exponential speed-up of quantum computers over classical ones in performing physical simulations \cite{Lloyd96} and solving hard mathematical problems like integer factorization \cite{Shor94}. However, implementing a quantum computer is difficult due to the technical obstacle of conveniently manipulating quantum degrees of freedom while preventing environmentally induced decoherence \cite{Unruh95Chuang95} at the same time. Solid-state quantum devices such as quantum dots \cite{QuantumDots} and SQUIDs \cite{SQUIDs} can be conveniently fabricated and are good at quantum state manipulation, but they are apparently very poor at preserving quantum coherence. All serious implementations of quantum logic exploit natural isolation in some systems such as certain cold trapped ions \cite{TrappedIons} which may stay at the metastable states for a long time, and nuclear spins \cite{NMRQC} that are well isolated from electronic and vibration motions. Preserving quantum coherence is achieved at the cost of awkwardness in logic manipulation and system scaling. Up to now, the nuclear spin approach appears the most promising in compromising system isolation and logic manipulation. But several factors make it hard to do quantum logic with a large system. To name but lacking the ability of addressing spins individually, logic operations must be distinguished by different RF pulses. The finite band width of radio frequency limits the size of the computer. Besides, it is not easy to explore the details of spin-spin interactions among a large molecule. As already mentioned, solid-state quantum logic is appealing because of the great convenience in device fabrication and control, especially the well-established technology to construct complex integrated circuits. Furthermore, a solid-state quantum computer will be compatible with conventional electronic computers which may serve as its peripheral equipments. Nevertheless, there is the decoherence problem to be cleared before solid-state quantum computers come out of fancy. Here we discuss the possibility of preserving quantum coherence in local energy minima (LEM) of a cluster of interacting quantum devices and performing quantum logic among such clusters.

It needs at least two distinct states to store a qubit, the smallest unit of quantum information. A physical system with a nondegenerate ground state may store the logic $0$ in its ground state while the logic $1$ must go to a state of higher energy. Quantum coherence is immediately spoiled whenever the state of the qubit bearer uncontrollably jumps, so energy dissipation is the first element to eliminate in order to maintain quantum coherence. Although keeping the system away from energy excitations or lowering the environmental temperature will effectively prevent upward jumps where energy is gained from the environment, the scheme fails in holding downward jumps back since the system may spontaneously decays and loses its energy even in the vacuum \cite{SpontaneousDecay}. Different from cold trapped ions and nuclear spins, individual solid-state quantum devices are always strongly coupled to their substrates, let alone the vacuum field. It is unlikely for them to have a long-living metastable state since the necessary perfect symmetry is always broken by fabrication imperfections. However, a cluster of such devices combined together may have states of LEM which are very long-living. The notion of LEM often appears in the context of spin-glasses \cite{SpinGlasses} or similar frustrated systems where it refers to a state well decoupled to lower energy states in the sense that all downward transitions are forbidden, spontaneous relaxation stops there. If an system is prepared in a superstition of the ground state and the LEM, dissipation is avoided and it becomes possible to keep quantum coherence for some time.

A single electron hopping between two weakly coupled quantum dots can be modeled by a pseudo-spin associated with two Pauli matrixes, $\sigma ^z$ and $\sigma ^x$. With the basis spanned by the two on-site states $(1\hspace{3mm}0)^T$ and $(0\hspace{3mm}1)^T$ where the electron is localized in one of the dots \cite{Tougaw93}, $\sigma ^z$ is diagonal, $\sigma ^z=\left(\begin{array}{cc}1&0\\ 0&-1\end{array}\right)$, while $\sigma ^x=\left(\begin{array}{cc}0&1\\ 1&0\end{array}\right)$ describes tunneling between the two sites. An RF SQUID is an isomorphic system characterized by a double well potential \cite{SQUIDs} thus falls into the same pseudo-spin model. Such a single pseudo-spin would not be very useful in storing quantum information because its strong coupling to the environment destroys quantum coherence rapidly. But things may be different in a cluster of pseudo-spins well-separated so that no tunneling among them but there are ``spin-spin'' interactions $A_{ij}\sigma ^z_i\sigma ^z_{j}$ ($i\neq j$) between two pseudo-spins which may originate from the Coulomb on-site repulsion in quantum-dot-clusters \cite{Tougaw93} or the magnetic coupling between trapped fluxes \cite{Davidovic96} in RF SQUIDs. Despite its various origin, the interaction is analogous to the Heisenberg exchange picture of ferromagnetic atoms. Indeed, in the context of molecular electronics \cite{Carter82and88}, one may envision doing quantum logic with real electronic spins in an elaborately designed magnetic molecule. Again, though individual spins may be coupled to the molecular vibrations, a spin-cluster could still serve as a good qubit bearer.

To understand why a cluster can stay at the LEM for a long time, let's simply consider a cluster of pseudo-spins (or even real spins) with ferromagnetic interactions $A_{ij}<0$,
\begin{equation}
{\bf H}=\sum_{i\neq j}
A_{ij}\sigma ^z_i\sigma ^z_{j}
+\sum_{i=1}^{n}B_i\sigma ^z_i
+\sum_{i=1}^{n}C_i\sigma ^x_i
\end{equation}
where the bias term $B_i\sigma ^z_i$ takes into account the possible difference of on-site energy of the $i$th pseudo-spin, and $C_i\sigma ^x_i$ cares the tunneling between the on-site states. Although the following discussion is based on this model Hamiltonian, the conclusion is valid for a wide variety of spin- or pseudo-spin-clusters. In the limit of very weak tunneling, $C_i\rightarrow 0$, ${\bf H}$ is diagonal in the basis expanded by $2^n$ binary ``number states'' $|x_1x_2\cdots x_n\rangle$, where $x_i = 1$ or $0$ corresponding to the up or down state of the $i$th pseudo-spin,
$$\sigma ^z_i|x_1x_2\cdots x_n\rangle =
\left\{
\begin{array}{rcl}
|x_1x_2\cdots x_n\rangle &,& x_i=1; \\
-|x_1x_2\cdots x_n\rangle &,& x_i=0.
\end{array}
\right.$$
Define the distance between two number states $|X\rangle =|x_1x_2\cdots x_n\rangle$ and $|Y\rangle=|y_1y_2\cdots y_n\rangle$ as $D(X,Y)=\sum _{i=1}^{n}(x_i-y_i)^2$. In case the exchange interaction is sufficiently larger than the bias, the two farthest states $|00\cdots 0\rangle$ and $|11\cdots 1\rangle$ have the lowest energies, with one, {\it e.g.}, $|11\cdots 1\rangle$ the LEM and the other the real ground state. Although an cluster at an LEM has extra energy, it stays there quite stably at low enough temperature since a spontaneous transition to the real ground state needs all the $n$ spins to flip simultaneously which is highly improbable. To be more specific, any spontaneous transition of the cluster is due to its coupling to the zero-point oscillations of electromagnetic fields or lattice waves. The coupling Hamiltonian can be generally written as
\begin{equation}
H'(t)=\sum _iF_i(t)\sigma _i^z+\sum _iG_i(t)\sigma _i^x,\label{FandGcoupling}
\end{equation}
where $F_i(t)$ give fluctuations of the on-site energy and $G_i(t)$ cause deviations of tunneling strength from the mean value (assumed to be zero at this stage). The rate of transition from $|X\rangle$ to $|Y\rangle$ is given by
\begin{equation}
R(X,Y)\propto\left| \langle X|H'(t)|Y\rangle \right|^2.\label{RXY}
\end{equation}
The $F$ terms do not lead to transition since they are diagonal in the space of the binary number states.
In the first order perturbation, the operator $\sum _iG_i(t)\sigma _i^x$ can only flip one spin, the rate
$R_1(X,Y)\equiv 0$ when $D(X,Y)>1$. Higher order perturbations of $\sum _iG_i(t)\sigma _i^x$ can
flip more spins, may eventually cause the notorious spontaneous transition. But the rate $R_d$ of multi-photon or multi-phonon processes \cite{SpontaneousDecay} is very small, decreasing exponentially as the distance $d=D(X,Y)$ increases, $R_d(X,Y)\sim (G/A)^dR_0$, where $G$ is the typical coupling strength between the spin and the perturbation field, $A$ is the typical level spacing in the spin-cluster, and $R_0$ is a constant. In practice, however, $C_i$ are not zero, though small. The ground state and the LEM are no longer exactly (but still nearly) $|00\cdots 0\rangle$ and $|11\cdots 1\rangle$. Other number states will mix in. The ground state $|X'\rangle$ is almost $|X\rangle =|00\cdots 0\rangle$, with other number states $|Z\rangle$ mixing in by small amplitudes $\langle X'|Z\rangle$. Using the stationary perturbation theory \cite{Schiff68}, it is straightforward to show $\langle X'|Z\rangle\sim (C/A)^{D(X,Z)}$, where $C$ is the typical value of $C_i$. $\langle X'|Z\rangle$ becomes exponentially small as the distance $D(X,Z)$ increases. Similarly, the LEM $|Y'\rangle$ is almost $|Y\rangle =|11\cdots 1\rangle$ with exponentially diminishing mixtures. Now evaluate the overall transition rate by Eq.~\ref{RXY}, taking into account both the tunneling effect and possible multi-photon or multi-phonon processes, one gets
\begin{equation}
R(X',Y')\leq \left(\frac{{\rm max}(C,G)}{A}\right)^{n}R_0
\end{equation}
where $n=D(X,Y)$ is the total number of pseudo-spins. We have done computer simulation for small clusters consisting of several spins whose Hamiltonian  can be exactly solved. The simulating results are well consistent with the above perturbation theory approach. The conclusion is that increasing the size of the cluster may exponentially prolong the lifetime of the LEM which together with the ground state can store a bit of quantum information. With reasonable values of ${\rm max}(C,G)/A=0.001\sim 0.01$, a cluster of $4$ or $5$ spins can efficiently extend the lifetime of the LEM by up to $10$ orders of magnitude. We note that in certain ``natural'' spin-clusters like a magnetic Mn$_{\rm 12}$ molecule, there is indeed a very stable LEM whose life-time could be years long at low temperature \cite{SpinCluster}. Such natural spin-clusters, if properly exploited, may serve as good quantum registers.

However, eliminating dissipation only prevents ``hard'' losses of quantum information accompanied by energy exchange, manifested as changes in the diagonal elements of the density matrix of the quantum system. The ``softer'' but more difficult to prevent is the loss of quantum coherence due to phase walk-off among the orthogonal states in a quantum superposition, manifested as the off-diagonal elements of the density matrix being quickly decreasing and vanishing, while the diagonal elements stay the same. In a model system of decoherence due to dephasing \cite{Omnes94}, a qubit system such as one made of a cluster of pseudo-spins is assumed to be at a typical superposition state $c_0|\psi_0\rangle+c_1|\psi_1\rangle$ at time $t=0$, $c_0,c_1\in{\bf C}$, $|c_0|^2+|c_1|^2=1$, which is in an environment consisting of a large number $N$ of two-state oscillators and at an initial state $\bigotimes_{n=1}^N\left(a_n|0_n\rangle+b_n|1_n\rangle\right)$, $a_n,b_n\in{\bf C}$, $|a_n|^2+|b_n|^2=1$, $\forall~n\in[1,N]$. It is further assumed that the environment is at a sufficiently low temperature comparing to the energy scale of the pseudo-spins so that no environmental oscillator is able to flip a single pseudo-spin in a cluster and exchange energy with it. Therefore, all interactions between the qubit and the environmental oscillators are diagonal terms that do not exchange spins and energy, $H_{\rm int}=\sum_{n=1}^N\hbar g_n\sigma^z\sigma^z_n$, where $\sigma^z$ and $\{\sigma^z_n\}_{n=1}^N$ are the diagonal Pauli matrices for the qubit and environmental oscillators, and $\{g_n\}_{n=1}^N\subset{\bf R}_+^N$ are coupling coefficients. If $2\hbar\omega$ and $\{2\hbar\omega_n\}_{n=1}^N$ are the two-level energy spacings, then the total Hamiltonian of the combined system is
\begin{equation}
H=\hbar\omega\sigma^z+\sum_{n=1}^N\hbar\omega_n\sigma^z_n+\sum_{n=1}^N\hbar g_n\sigma^z\sigma^z_n,
\end{equation}
which governs the evolution of the combined system in time. The quantum state of the whole system is exactly solvable and reads \cite{Omnes94}
\begin{eqnarray}
|\Psi(t)\rangle&=&c_0e^{i\omega t}|\psi_0\rangle\bigotimes_{n=1}^N
\left(a_ne^{i\omega_nt}e^{-ig_nt}|0_n\rangle\right.\nonumber\\
&&~~~~~~~~~~~~~~~~+\left.b_ne^{-i\omega_nt}e^{ig_nt}|1_n\rangle\right)
\nonumber\\
&+&c_1e^{-i\omega t}|\psi_1\rangle\bigotimes_{n=1}^N
\left(a_ne^{i\omega_nt}e^{ig_nt}|0_n\rangle\right.\nonumber\\
&&~~~~~~~~~~~~~~+\left.b_ne^{-i\omega_nt}e^{-ig_nt}|1_n\rangle\right).
\end{eqnarray}
Although the interactions between the qubit and the environmental oscillators do not involve energy exchange, it is clear that they lead to conditional phase shifts, which may be interpreted as that each environmental oscillator undergoes ``controlled'' phase shifts between the ``0'' and ``1'' states, depending on the state of the qubit. Such conditional phase shifts, if randomly distributed and uncorrected, leads to decoherence. $\forall~t\ge 0$, the density matrix of the combined system is $\rho(t)=|\Psi(t)\rangle\langle\Psi(t)|$. As it is impossible to monitor and control the environmental degrees of freedom, one chooses to trace over the environmental degrees of freedom and obtain a reduced density matrix for the qubit system \cite{Omnes94},
\begin{eqnarray}
\rho_Q(t)&=&\sum_{k_1=0}^1\cdots\sum_{k_N=0}^1\langle k_1|\cdots\langle k_N|\rho|k_N\rangle\cdots|k_1\rangle\nonumber\\
&=&|c_0|^2|\psi_0\rangle\langle\psi_0|+|c_1|^2|\psi_1\rangle\langle\psi_1|\nonumber\\
&+&z(t)c_0c^*_1e^{i2\omega t}|\psi_0\rangle\langle\psi_1|\nonumber\\
&+&z^*(t)c^*_0c_1e^{-i2\omega t}|\psi_1\rangle\langle\psi_0|,
\end{eqnarray}
with the factor
\begin{equation}
z(t)=\prod_{n=1}^N\left[\cos 2g_nt+i\left(|b_n|^2\!-\!|a_n|^2\right)\sin 2g_nt\right].
\end{equation}
It is easy to see that $z(0)=1$, but for even a small $t=\epsilon>0$, the modulus $|z(t)|$ becomes
\begin{eqnarray}
&&\prod_{n=1}^N\left|1\!-\!2g_n^2\epsilon^2\!+\!i\left(|b_n|^2\!-\!|a_n|^2\right)
\!2g_n\epsilon\!+\!O(\epsilon^3)\right|\nonumber\\
&=&\!\prod_{n=1}^N\left\{1\!-\!\left[1\!-\!\left(|b_n|^2\!-\!|a_n|^2\right)^2\right]
\!2g_n^2\epsilon^2\!+\!O(\epsilon^3)\right\},
\end{eqnarray}
which quickly diminishes in the limit $N\rightarrow\infty$, provided that $\{(a_n,b_n)\}_n$ are randomly distributed and most of the coupling coefficients $\{g_n\}_n$ are not vanishingly small. In that case, the reduced density matrix quickly becomes diagonal, representing a classical mixed state with little to no quantum coherence.

It is cleat that in order to prevent fast decoherence or elongate the quantum coherence time, both the number of environment oscillators coupled to the qubit and their coupling coefficients $\{g_n\}_n$ need to be minimized. Keeping the environment at low temperature helps to reduce the number of, or the density of states of, environmental oscillators that can be excited to nontrivial states $a_n|0_n\rangle+b_n|1_n\rangle$, $n\in{\bf N}$. It is advantageous to neutralize and symmetrize the distribution of charge and electric or magnetic moments of the qubit states, so that the lower-order multipole moments in the multipole expansion \cite{Jackson75} vanish, and the possible electric or magnetic interactions between the qubits and environmental oscillators are highly localized. More specifically, if a qubit state has multipole moments canceled up to the $m$th order, $m\ge 0$, then the potential field of all the non-vanishing multipole moments (of orders $\ge(m+1)$) decays at least as fast as $r^{-(m+2)}$ in the distance $r$ away from the center of the pseudo-spin cluster of the qubit \cite{Jackson75}. The length scale of the multipole expansion and field decay is determined by the physical size of the pseudo-spin cluster, which should be minimized. For environmental oscillators that are localized and particles in nature, such as pseudo-spins of other qubits, impurity atoms, pinned electric dipoles and magnetic moments, the couplings with a qubit reduces quickly to negligible as they get further away, only those that are in close proximity may be coupled to the qubit. For environmental oscillators that are extensive and waves in nature, such as electromagnetic radiations and crystal lattice vibrations, the couplings average toward zero for modes with wavelengths that are significantly longer than the physical size of the pseudo-spin clusters of the qubits, only shorter-wavelength (thus higher-frequency) modes may experience a net coupling. Therefore, reducing the density of impurities, lowering the environment temperature, symmetrizing the pseudo-spin clusters, and minimizing the physical size of pseudo-spin clusters are general rules and means to reduce the number of effective environmental oscillators and minimizing their couplings to qubits, so to keep quantum coherence for an appreciable duration of time. Here it may be worth noting that the later development of ``topological quantum computation'' \cite{TopologicalQC} shares a bit of the same spirit in that quantum information is stored in global multi-body states or ``noise-free subspaces'' which are immune to local, {\it i.e.}, single- or few-body perturbations.

Apart from such inherent robustness against decoherence by physical designs, there are also algorithmic measures, including quantum error correction coding \cite{ShorSteane,NewScientist} and dynamical decoupling \cite{DynamicalDecoupling}, to protect the delicate quantum coherence against unavoidable noise and decoherence. A quantum information processor in the future may employ both measures of physical designs and error correction algorithms complementarily, where judicious designs of devices ensure a raw rate of quantum error below a certain threshold, so that error correction algorithms are able to overcome decoherence and may be iterated to allow an indefinite number of reliable quantum logic operations. Finally, to perform quantum logic operations, it would be necessary to bring or place pseudo-spin clusters close to interact, and to apply external excitations to drive the qubits switching between the ``0'' and ``1'' states depending on the states of their neighbors, for which purpose, the ground and LEM states of a pseudo-spin cluster may need to be coupled to an intermediate excited state, exposing the quantum information to a hazardous environment. This may cause ``gate errors''. Fortunately, just like ``memory errors'', ``gate errors'' can also be corrected \cite{Cirac96}. And as in the context of quantum optics, the technique of ``dark-state transfer'' \cite{Bennett95,Scully97} should effectively avoid populating spin-clusters to excited states. Moreover, it is even possible to do quantum computation directly on encoded qubits so that quantum coherence is always under protection \cite{NewScientist,ZurekEtAl}.

In summary, solid-state quantum devices may be designed and optimized for storing and processing quantum information, where pseudo-spin clusters are symmetrized to lower the rate of decoherence. Once the problem of decoherence is overcome and quantum error correction is made practical, the excellent architectural flexibility and ease of scaling as well as compatibility with classical electronics as peripherals are appealing promises held by solid-state quantum devices in the emerging field of quantum information processing.

\end{document}